\documentclass[aps,superscriptaddress,twoside,twocolumn,floatfix,a4paper,longbibliography,prl]{revtex4}

\usepackage{graphicx}
\usepackage{amsmath}
\usepackage{dcolumn}   
\usepackage{bm}        
\usepackage{amssymb}   
\usepackage{graphicx,epsfig}
\usepackage{color}
\usepackage[usenames,dvipsnames]{xcolor}
\usepackage[ocgcolorlinks,colorlinks=true,linkcolor=blue,citecolor=red]{hyperref}
\usepackage{amsmath,amssymb, amsthm, amsfonts}
\usepackage{xspace}
\usepackage{xcolor}
\usepackage{verbatim}
\usepackage{mathtools}

\begin{document}

\title{Field-Effect Controllable Metallic Josephson Interferometer}

\author{Federico Paolucci}
\email{federico.paolucci@nano.cnr.it}
\affiliation{ INFN Sezione di Pisa, Largo Bruno Pontecorvo, 3, I-56127 Pisa, Italy}
\affiliation{ NEST, Instituto Nanoscienze-CNR and Scuola Normale Superiore, I-56127 Pisa, Italy}
\affiliation{These authors equally contributed to this work.}
\author{Francesco Vischi} 
\affiliation{ NEST, Instituto Nanoscienze-CNR and Scuola Normale Superiore, I-56127 Pisa, Italy}
\affiliation{ Dipartimento di Fisica dell'Universit\`a di Pisa, Largo Pontecorvo 3, I-56127 Pisa, Italy}
\affiliation{These authors equally contributed to this work.}
\author{Giorgio De Simoni}
\affiliation{ NEST, Instituto Nanoscienze-CNR and Scuola Normale Superiore, I-56127 Pisa, Italy}
\author{Claudio Guarcello}
\affiliation{ NEST, Instituto Nanoscienze-CNR and Scuola Normale Superiore, I-56127 Pisa, Italy}
\author{Paolo Solinas}
\affiliation{ SPIN-CNR, Via Dodecaneso 33, I-16146 Genova, Italy}
\author{Francesco Giazotto}
\email{francesco.giazotto@sns.it}
\affiliation{ NEST, Instituto Nanoscienze-CNR and Scuola Normale Superiore, I-56127 Pisa, Italy}



\begin{abstract}
Gate-tunable Josephson junctions (JJs) are the backbone of superconducting classical and quantum computation. Typically, these systems exploit low charge concentration materials, and present technological difficulties limiting their scalability. Surprisingly, electric field modulation of supercurrent in metallic wires and JJs has been recently demonstrated.
Here, we report the realization of titanium-based monolithic interferometers which allow tuning both JJs independently via voltage bias applied to capacitively-coupled electrodes.
Our experiments demonstrate full control of the amplitude of the switching current ($I_S$) and of the superconducting phase across the single JJ in a wide range of temperatures.
Astoundingly, by gate-biasing a single junction the maximum achievable total $I_S$ suppresses down to values much lower than the critical current of a single JJ.
A theoretical model including gate-induced phase fluctuations on a single junction accounts for our experimental findings. This class of quantum interferometers could represent a breakthrough for several applications such as digital electronics, quantum computing, sensitive magnetometry and single-photon detection.
\end{abstract}


\maketitle

The possibility of tuning the properties of Bardeen-Cooper-Schrieffer (BCS) metallic superconductors via conventional gating has been excluded for almost a century.
Surprisingly, strong static electric fields have been recently shown to modulate the supercurrent down to full suppression and even to induce a superconductor-to-normal phase transition in metallic wires \cite{DeSimoni2018, Varnava2018} and Josephson junctions (JJs) \cite{Paolucci2018, Paolucci2019, DeSimoni2019} without affecting their normal-state behavior.
Yet, these results did not find a microscopic theoretical explanation so far \cite{Virtanen2019}.
In this Article, we lay down a fundamental brick for both the insight and the technological application of this unorthodox field-effect by realizing a titanium-based monolithic superconducting quantum interference device (SQUID) which can be tuned by applying a gate bias to both JJs independently.
We first show modulation of the amplitude and the position of the interference pattern of the switching current ($I_S$) by acting with an external electric field on a \textit{single} junction of the interferometer.
Notably, this phenomenology differs from that of conventional gate-tunable SQUIDs \cite{vanDam2006,Cleuziou2006,Girit2008,Goswami2016} and cannot be explained by a simple squeezing of the critical current of the junction induced by the electric field \cite{Clarke2004}.
Consequently, a local electric field acts on a global scale influencing the properties of both JJs.
Since the superconducting phases of the two JJs are non-locally connected by fluxoid quantization, we deduce that the electric field must act both on the critical current amplitude and \emph{couple} to the superconducting phase across the single junction.
The overall interferometer phase can shift from $-0.4\pi$ to $0.2\pi$ depending on the used gate electrode and on the strength of the gate bias.
Furthermore, the effect persists up to $\sim80\%$ of the superconducting critical temperature.
Fully-metallic field-effect controllable Josephson interferometers could lay the first stone of novel superconducting architectures suitable for classical \cite{Clark1980, Terzioglu1998} and quantum \cite{Devoret2013, Majer2002, Ioffe1999, Mooij1999} computing, and for ultrasensitive tunable magnetometry \cite{Anahory2014, Uri2016}.

\begin{figure*} [ht!]
\begin{center}
\includegraphics{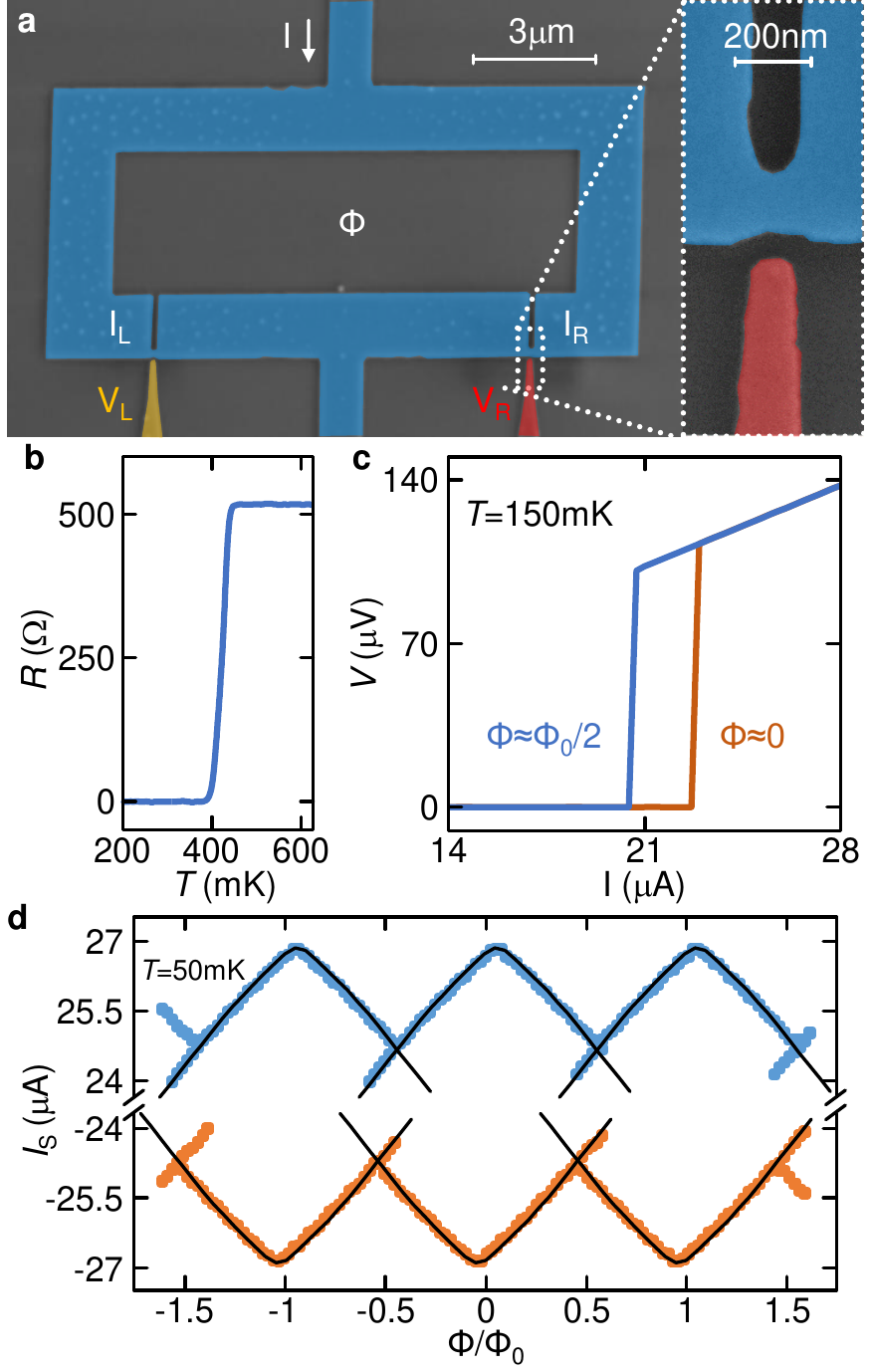}
\end{center}
\caption{\label{fig:Fig1}\textbf{Interferometer layout and basic characterization.} \textbf{a} False-color electron micrograph of a typical all-metallic field-effect controllable Josephson interferometer. The core of the SQUID is represented in blue whereas the left side gate ($V_L$) in yellow and the right side gate ($V_R$) in red. $\Phi$ denotes the magnetic flux piercing the loop while $I$ is the current flowing through the interferometer. The inset shows a blow-up of the right gate-tunable Josephson junction. \textbf{b} Four-probe resistance $R$ versus temperature $T$ characteristic of the entire device. \textbf{c} Individual voltage $V$ versus current $I$ traces measured at $T=150$ mK for $\Phi\simeq0$ (orange) and $\Phi\simeq\Phi_0/2$ (blue). \textbf{d} Positive (blue squares) and negative (orange squares) switching currents $I_S$ of the SQUID measured at $T=50$ mK. The black lines represent the theoretical prediction for a Dayem-bridge-based SQUID calculated within the RCSJ formalism (see Methods for further details). The resulting error bars are smaller than the line dimension.}
\end{figure*}

Figure \ref{fig:Fig1}-a shows a typical field-effect controllable metallic Josephson interferometer. The monolithic dc-SQUID is realized in the form of a 30-nm-thick Ti superconducting loop interrupted by two 150-nm-long and 150-nm-wide nano-constrictions (Dayem-bridges), that we label as $L$ for left and $R$ for right. Corresponding to each constriction a side electrode ($V_L$ and $V_R$, at a distance of about 30 nm and 50 nm, respectively) allows to independently control the switching current of the JJ ($I_{L}$ and $I_{R}$) via an electrostatic field \cite{DeSimoni2018, Paolucci2018, Paolucci2019}. The details of the simple single-step fabrication process can be found in the Methods.

We start the basic investigation of our interferometer by measuring its resistance $R$ versus temperature $T$ characteristic shown in Fig. \ref{fig:Fig1}-b. In particular, a critical temperature $T_C\simeq420$ mK and a normal-state resistance of the entire device $R_N\simeq550$ $\Omega$ are recorded.  To study the current-flux behavior of the SQUID, we measured the voltage versus current characteristics for $\Phi\simeq0$ (orange) and $\Phi\simeq\Phi_0/2$ (blue), where $\Phi _0\simeq 2\times 10^{-15}\;\textup{Wb}$ is the flux quantum, at $T=150$ mK (see Fig. \ref{fig:Fig1}-c). Denoting with $I_S(\Phi)$ and $I_{S,ave}$ the switching current as a function of the magnetic flux and the average current, respectively, a modulation visibility $\Delta I_S /I_{S,ave}\simeq11\%$ is observed, where $\Delta I_S=I_S(0)-I_S(\Phi_0/2)$. The basic parameters of the titanium thin film and the SQUID extracted from the experimental data are presented in the Methods section.

The characteristic triangular magnetic-flux patterns \cite{Fulton1972, Tsang1975} of both positive (blue) and negative (orange) branch of the switching current obtained at $T=50$ mK are represented in Fig. \ref{fig:Fig1}-d. In order to extract the basic parameters of the interferometer we developed a theoretical model based on the resistively and capacitively shunted junction (RCSJ) formalism (see Methods for further details). The black lines in Fig. \ref{fig:Fig1}-d are obtained by considering a working temperature of $50$ mK and the presence of thermal fluctuations on both JJs ($I_L^{\text{th}}$ and $I_R^{\text{th}}$). We deduced the screening parameter, $\beta \simeq32$, accounting for the multiple value of $I_S$ for certain values of magnetic flux \cite{Russo2014}, and a small asymmetry in the critical current of the two junctions ($I_{C,L}\simeq13.6$ $\mu$A and $I_{C,R}\simeq13.3$ $\mu$A, respectively) causing the limited offset of $I_S(\Phi)$ along the external magnetic flux axis ($\simeq 0.05 \Phi_0$). We would like to stress that in our experimental conditions the thermal fluctuations give a negligible contribution to the phase dynamics of the SQUID, that is the error bars of the black curve of Fig. \ref{fig:Fig1}-d are vanishingly small.

\begin{figure*} [t!]
\begin{center}
\includegraphics [width=0.9\textwidth]{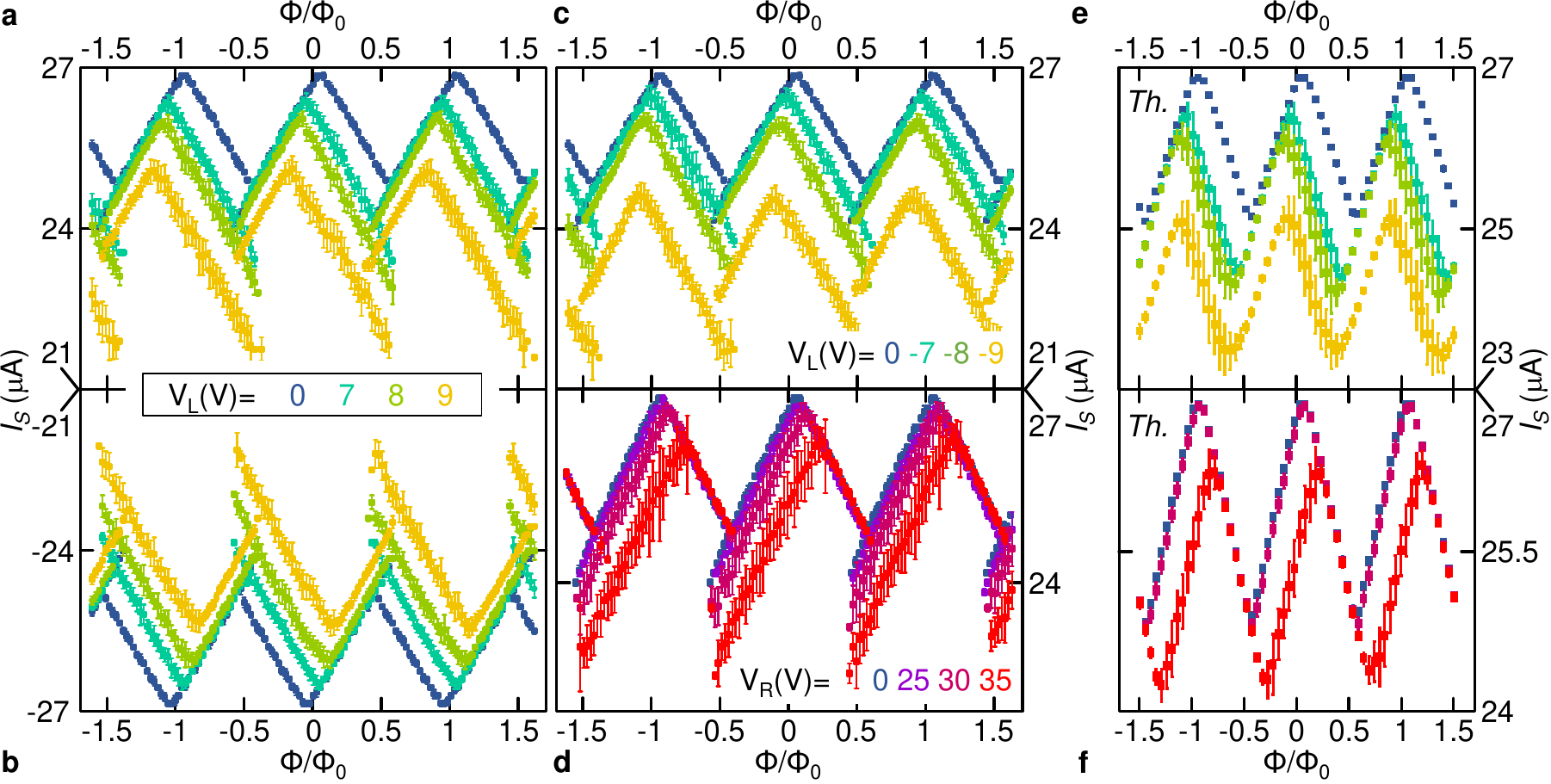}
\end{center}
\caption{\label{fig:Fig2} \textbf{Electric field-dependent magnetic modulations: experiments and theoretical model.} Experimental modulations of the switching current $I_S$ with the magnetic flux $\Phi$ piercing the SQUID at $T=50$ mK for positive (\textbf{a}) and negative (\textbf{b}) current bias measured for positive values of voltage applied to the left gate electrode $V_L$. (\textbf{c} ) Positive branch of $I_S(\Phi)$ for different negative values of $V_L$  and for different values of positive right gate bias $V_R$ (\textbf{d}) measured at $T=50$ mK.
In the experiment, the leakage current measured between the gate electrode and the SQUID is typically of the order of a few tens of pA resulting in a gate-SQUID impedance of a few T$\Omega$. Theoretical modulations of $I_S$ with $\Phi$ calculated with our RCSJ-based model for different values of current fluctuations and critical current suppression applied to the left (\textbf{e}) and right (\textbf{f}) JJ. The parameters used in the numerical simulations are listed in the Supplementary Information.}
\end{figure*}

To ensure independent field-effect control of a single JJ, the intensity of the electric field generated through the left (right) gate electrode at the surface of the right (left) Dayem bridge has to be negligible. To this end, we fabricated the junctions separated by roughly $8~\mu$m (see Fig. \ref{fig:Fig1}-a). Furthermore, finite element simulations performed for the same geometry of our devices showed that the resulting  $\| \vec{D} \|$ is suppressed by about a factor of 20 at the right JJ so that the effect of $V_L$ on $I_R$ can be considered negligible (see Supplementary Information for further details).

To investigate the impact of the gate bias on the SQUID switching current characteristic, we measured $I_S(\Phi)$ for different values of gate voltage independently applied to both the left ($V_L$) and right ($V_R$) gate electrode.
Figures \ref{fig:Fig2}-a and b show the magnetic modulation patterns of the positive and negative branch of the switching current for different positive values of $V_L$ measured at $T=50$ mK. The right gate electrode has been left floating or grounded with no evident change in the SQUID behavior.
The data highlight several interesting gate-dependent features.
First, the maximum switching current $I_{S,max}$ is suppressed by increasing gate voltage, and it shows a similar reduction for positive and negative current bias.
Second, $I_{S,max}$ shifts towards lower (higher) magnetic flux values for the positive (negative) branch of the switching current.
Third, the experimental error bars of $I_S$ are different in the first and second half of the oscillation period with the magnetic flux.
They grow for the first half period of the switching current, whereas they remain almost constant for the second one.

\begin{figure} [t!]
\begin{center}
\includegraphics {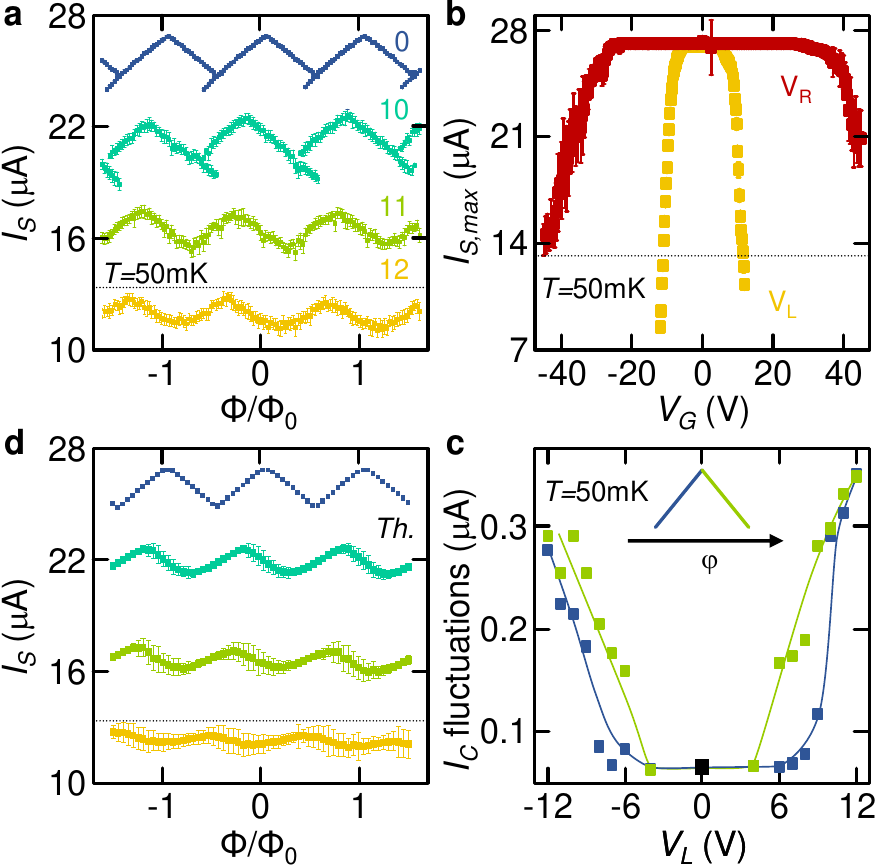}
\end{center}
\caption{\label{fig:Fig3} \textbf{Impact of electric field on the superconducting phase.} \textbf{a} $I_S(\Phi)$ for different values of gate voltage $V_L$ measured at $T=50$ mK. The dotted black curve represents the intrinsic value of the critical current of the right JJ ($I_{C,R}\simeq13.3~\mu$A).
\textbf{b} Switching current for zero magnetic flux piercing the interferometer $I_{S,max}$ as a function of the bias voltage applied to left (yellow squares) and right (red squares) gate electrode at $T=50$ mK.
\textbf{c} Average fluctuations of SQUID switching current as a function of the bias applied to the left gate electrode $V_L$ for the rising (blue squares) and decreasing (green squares) branch at $T=50$ mK.
\textbf{d} Switching current $I_S$ as a function of the external magnetic flux $\Phi$ calculated within an \emph{ad hoc} model based on the RCSJ formalism. In particular, we considered the joint impact of the dampening of the critical current $I_{C,L}$ of the left nano-constriction and the enhancement of the electric field-dependent superconducting current fluctuations $I_L^{\text{gate}}$ on the same JJ. The parameters used in the numerical simulations are listed in the Supplementary Information.}
\end{figure}

Importantly, field-effect is almost symmetric in the polarity of $V_L$ (see Fig. \ref{fig:Fig2}-c), as reported for the reduction of $I_S$ in metallic wires \cite{DeSimoni2018} and Dayem-bridge JJs \cite{Paolucci2018, Paolucci2019}. As a consequence, any charge accumulation or depletion seem to be excluded since they would provide an opposite effect on the magnitude of the switching current. Furthermore, in order to exclude any quasiparticle overheating due to direct current injection from the gate to the JJs, during all experiments we monitored the leakage current $I_L$. The leakage current is always on the order of tens of picoamps for both gate electrodes and it is not correlated with the strength of the $I_S$ suppression (see Supplementary Information for further details). In addition, field-effect control of titanium Dayem bridges has been demonstrated to be unaffected from leakage currents of this order of magnitude \cite{Paolucci2018,Paolucci2019}.

In full agreement with the data obtained for $V_L$, by applying a bias $V_R$ to the right gate electrode yields suppression of the maximum switching current (see Fig. \ref{fig:Fig2}-d).
Yet, the magnetic modulation patterns shift towards higher values of the magnetic flux and the larger current fluctuations are recorded for the second half period of $I_S(\Phi)$ oscillation.
In addition, the values of $V_R$ which are necessary to affect $I_S$ are bigger than for $V_L$ because the distance between the JJ and the gate electrode is larger in $R$ than in $L$.

Previous gating experiments performed on metallic JJs \cite{Paolucci2018, Paolucci2019, DeSimoni2019} suggest that the electric field can strongly reduce their switching current. We included this information in the model by reducing $I_{C,L}$ or $I_{C,R}$ (with $I_{C,L}$ and $I_{C,R}$ denoting the switching currents of the left and right JJ, respectively) when the corresponding gate bias is applied.
In this case, a sizable screening parameter $\beta$ is enough to explain both the reduction of the total switching current of the SQUID and the shift of the interference pattern along the magnetic flux axis \cite{Clarke2004, Bar82}. However, this is not sufficient to account for the difference in the switching current errors for the first and second half period of $I_S$ (see Fig. \ref{fig:Fig2}-a).
To this end, we introduced two gate-dependent fluctuating currents, $I_L^{\text{gate}}$ and $I_R^{\text{gate}}$, that disturb the dynamics of the corresponding superconducting phases.
This will be a key point to explain some of the features of the following measurements.

Figures \ref{fig:Fig2}-e and f show the result of the calculation of the joint impact of the switching current reduction and phase fluctuations in the left and right JJ on the complete interference pattern of the SQUID, respectively. By imposing the phase noise on the left (right) JJ, the interference pattern shifts towards negative (positive) values of the magnetic flux, and the $I_S$ fluctuations grow in the first (second) half of the switching current oscillation period.
The theoretical curves display a good qualitative agreement with the experimental data.
In particular, the calculations suggest that current fluctuations occur already for low values of gate voltage, while the critical current suppression start to develop for higher gate biases.  As a consequence, our data seem to highlight that an external static electric field can affect, even if indirectly, the phase of a metallic conventional superconductor.

Next step is to investigate the impact of the electrostatic field on the superconducting phase for higher values of gate voltage. Figure \ref{fig:Fig3}-a shows the experimental traces measured for $V_L\ge10$ V at $T=50$ mK.
The most prominent feature is the suppression of $I_S$ below the critical current of a single JJ for every value of the magnetic flux [$I_{S}(V_L=12\text{V})\le I_{C,R}$, see black dotted line in Fig. \ref{fig:Fig3}-a], while the oscillatory behavior is preserved with an almost constant visibility.
The stark reduction of the switching current is highlighted by plotting $I_{S,max}=I_S(\Phi=0)$ as a function of gate voltage (see Figure \ref{fig:Fig3}-b).
In particular, for $V_L=-12$ V the switching current surprisingly lowers down to $\sim0.6 I_{C,R}$. Furthermore, the asymmetry in the switching current fluctuations for the first/second half of the oscillation period of $I_S(\Phi)$ tends to disappear for high values of gate bias (see Figure \ref{fig:Fig3}-c).
In addition, the multivalued nature of $I_S$ for certain values of $\Phi$ turns out to vanish by increasing gate voltage.

\begin{figure*} [t!]
\begin{center}
\includegraphics [width=0.9\textwidth]{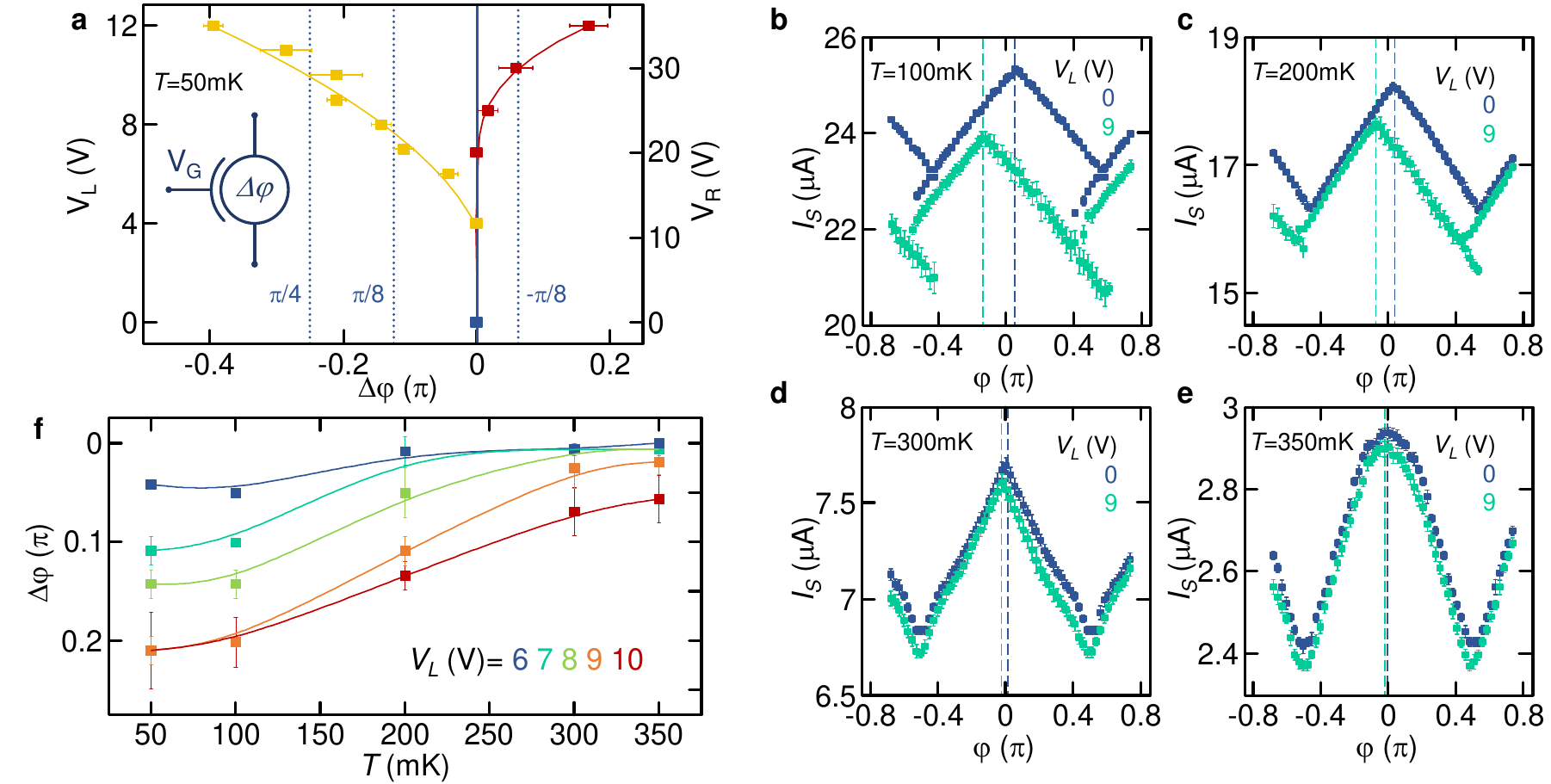}
\end{center}
\caption{\label{fig:Fig4}  \textbf{Phase shift and its temperature dependence.} \textbf{a} Phase shift $\Delta\varphi$ as a function of the voltage
 applied  to left (yellow squares) and right (red squares) gate electrode at $T=50$ mK. Dashed lines indicate typical phase shifts required for implementing quantum logic gates: $\pi/8$ and $\pi/4$ (the so-called $T$ gate).
The inset shows the schematic representation of a gate-tunable phase shifter.
\textbf{b-e} Modulations of the switching current $I_C$ with the superconducting phase $\varphi$ measured at four different temperatures $T=100, 200, 300, 350$ mK for two values of $V_L$ (0, blue squares, and 9 V, turquoise squares).
The dashed lines represent the phase position of $I_{C,max}$ for $V_L=0$ (blue) and $V_L=9$ V (turquoise).
\textbf{f} Phase shift $\Delta\varphi$ as a function of temperature for different values of the voltage applied to the left gate electrode $V_L$. The squares depict the averages of $\Delta\phi$ extracted from different periods with the associated error, while the solid lines are guides for the  eye.}
\end{figure*}

All the above phenomenology does not find any satisfactory explanation within conventional models for a SQUID \cite{Clarke2004, Bar82} with the assumption that the junction critical current decreases under the effect of the electric field \cite{DeSimoni2018, Paolucci2018, Paolucci2019, DeSimoni2019} or for a possible direct hot electron injection \cite{Morpurgo1998}.
In fact, if a single JJ is squeezed up to closure, a continuous and flux-independent flow of current in the ungated junction is expected to occur \cite{Clarke2004}.
By contrast, Figs. \ref{fig:Fig3}-a and b imply that a local static electric field influences both junctions thereby affecting the whole interferometer.

Figure \ref{fig:Fig3}-d shows the evolution of $I_S(\Phi)$ calculated by imposing, on the one hand, the dampening of the critical current $I_{C,L}$ of the left JJ and, on the other hand, the strengthening of the gate-dependent current fluctuations on the same junction $I_L^{\text{gate}}$.
These are \textit{local} noise sources exclusively acting on the left JJ due to the presence of the applied gate voltage ($V_L$).
Yet, in the SQUID configuration the superconducting phases are locked by fluxoid quantization \cite{Tinkham}. Therefore, any local perturbation affects non-locally the dynamics of the opposite junction, and all the \textit{global} observables of the interferometer.
The theoretical curves exhibit a good qualitative agreement with the experiment showing the suppression of $I_S$ below the value of the ungated junction (Fig. \ref{fig:Fig3}-d).
In addition, they are able to grab the evolution from a triangular-shaped interference pattern of $I_S(\Phi)$ to a smoother oscillatory behavior. We can therefore conclude that an external electrostatic field \textit{couples} with the superconducting phase of a conventional metallic superconductor.

The sliding of the interference pattern along the magnetic flux axis driven by the gate bias can be inherently projected in a tunable shift of the superconducting phase \cite{Clarke2004}. We define the phase shift as $\Delta\varphi(V_i)=\varphi(V_i)-\varphi(0)$, where $V_i$ is the voltage applied to the $i=L,R$ gate electrode. Figure \ref{fig:Fig4}-a shows the phase shift as a function of $V_L$ (yellow squares) and $V_R$ (red squares) extrapolated from our experimental data obtained at $T=50$ mK. Notably, the total maximum phase shift is $\Delta\varphi_{tot}=\Delta\varphi_{max}(V_R)-\Delta\varphi_{max}(V_L)\simeq0.6\pi$, where $\Delta\varphi_{max}(V)$, with $V=V_R,V_L$, is the maximum phase shift generated by applying the voltage $V$. The asymmetry in the effect of the two gate voltages on $\Delta\varphi$ can be entirely ascribed to the difference in the distance of the two gate electrodes from the corresponding weak-link regions \cite{Paolucci2019}.
From our data we can extrapolate a possible total maximum phase shift of at least $0.8\pi$ within this geometry, since a $\Delta\varphi\sim0.4\pi$ was obtained by biasing the left gate electrode (see Fig. \ref{fig:Fig4}-a).
In addition, we wish to stress that the maximum achievable phase shift could rise by applying stronger gate biases until the oscillatory behavior of $I_S(\Phi)$ is preserved.

The dependence of the switching current interference pattern on temperature provides important information about the  impact of the electric field on the Josephson coupling and on the efficiency of the phase shifter.
In particular, Figs. \ref{fig:Fig4}-b-e show the evolution of the ungated $I_S(\varphi)$ traces (blue squares) with temperature. On the one hand, the transition from a linear to a smooth trace is related to the current-phase relation of the JJs approaching the conventional sinusoidal behavior by increasing the temperature \cite{Fulton1972, Tsang1975}.
On the other hand, the disappearance of a multivalued $I_S(\Phi)$ is related to the decrease of the screening parameter \cite{Russo2014}.
The latter, together with a lower impact of the electric field on the supercurrent \cite{DeSimoni2018, Paolucci2018, Paolucci2019}, is also reflected in the decrease of the phase shift caused by $V_L=9$ V at higher temperatures (see Figs. \ref{fig:Fig4}-b-e).

The full behavior of $\Delta\varphi$ on temperature for different values of gate bias applied to the left JJ is summarized in Fig. \ref{fig:Fig4}-f. The phase shift decreases monotonically by increasing $T$ for all the applied values of gate voltage.
However, the phase shift persists up to about $85\%$ of $T_C$ for $V_L\ge8$ V denoting a wide range of operating temperatures.

Beyond the strong implications in fundamental physics, i.e., the apparent coupling of a static electric field to the macroscopic phase of the superconducting condensate, gate-controllable metallic Josephson interferometers may be employed to develop several scalable superconducting field-effect devices with a wide range of possible applications.
For instance, the possibility to manipulate and change the interferometric phase can yield  important implications in quantum information \cite{Majer2002, Ioffe1999,Sleator1995,Lloyd1995} in the form of flux \cite{Mooij1999, Yan2016} and phase \cite{Martinis2002} superconducting qubits \cite{Devoret2013,Koch2007,Schreier2008,Barends2013}, and in superconducting electronics \cite{Terzioglu1998} for the realization of electrostatically-tunable rapid single flux quantum (RSFQ) logic \cite{Likharev1991,Worsham1995}.
Furthermore, the ability of electrostatically mastering $I_S(\Phi)$ \cite{Goswami2016} could be of fundamental importance for the implementation of ultrasensitive tunable magnetometers \cite{Anahory2014, Uri2016}.
Finally, these prototypical interferometers could find application in phase-controllable caloritronics \cite{Fornieri2017,Giazotto2014} and single photon sensing \cite{Virtanen2018}.

\section*{Methods}

\subsection*{Fabrication}
The samples were nano-fabricated by single step electron-beam lithography and evaporation of Ti through a poly(methyl methacrylate) (PMMA) mask onto an intrinsic Si wafer covered by $300$ nm of silicon dioxide. The 30 nm thick Ti film was deposited at room temperature in an ultrahigh-vacuum electron-beam evaporator with a base pressure of about $5\times10^{-11}$ torr at a rate of about $11$ \AA/s.

\subsection*{Measurements}
All measurements were performed in a filtered He$^3$-He$^4$ dry dilution refrigerator at different bath temperatures in the range 50mK - 450mK using standard four-wire technique.
The resistance versus temperature characteristics were obtained by low frequency lock-in technique. To this end, a.c. excitation currents with typical root mean square amplitudes $I\simeq10$ nA at a frequency of $13.33$ Hz were imposed through the device.
The magnetic-flux patterns of the devices were obtained by injecting a low-noise biasing current and measuring the voltage drop by a room-temperature differential preamplifier. The gate bias was applied by a low-noise high-input impedance source/measure unit or a DC-voltage source adding low frequency filters of time constant ranging from 1 second to 100 seconds (without any change of the resulting data). The perpendicular-to-plane magnetic field is applied using a standard commercial superconducting coil magnet with the sample placed in its center. Every measurement point is the average over 50 repetitions and the respective error is their standard deviation.

\subsection*{Parameters of the titanium thin film and SQUID}
From the resistance versus temperature measurements (see Fig. \ref {fig:Fig1}-b) and the dimensions of the device we can deduce all the basic parameters of the titanium film and the SQUID.

The zero-temperature BCS energy gap is $\Delta_0=1.764k_BT_C\simeq64$ $\mu$eV, where $k_B$ is the Boltzmann constant and $T_C\simeq420$ mK is the critical temperature. From the diffusion constant ($D\simeq10^{-3}$ m$^2$) we extracted the conductivity $\sigma=DN_fe^2\simeq3.5\times 10^6$ S, where $N_f\simeq1.35\times10^{47}$ J$^{-1}$m$^{-3}$ is the density of states at the Fermi level of titanium and $e$ is the electron charge. The resulting magnetic field London penetration depth is $\lambda_L=\sqrt{\hbar/\sigma\mu_0\Delta_0}\simeq900$ nm, where $\hbar$ is the reduced Planck constant and $\mu_0$ is the magnetic permeability of vacuum, while we obtained a superconducting coherence length $\xi=\sqrt{\hbar D/\Delta_0}\simeq100$ nm.

The total inductance of the SQUID ($\mathcal{L}$) is given by the sum of the geometric and kinetic contributions. The geometric inductance of the SQUID ring is calculated with the expression for a planar spiral of rectangular section \cite{Hurley1995,Mohan1999} and it is $\mathcal{L}_G\simeq22.5$ pH. The kinetic inductance of the SQUID ring takes the form \cite{Annunziata2010} $\mathcal{L}_K= \frac{R_Rh}{2\pi^2\Delta}\frac{1}{\tanh\left(\frac{\Delta}{2k_BT}\right)}\simeq0.6$ nH, where $R_R=200$ $\Omega$ is the normal-state resistance of the ring, $h$ is the Planck constant, $T=50$ mK is the temperature, and $\Delta=\Delta_0$ is the superconducting gap (since $T/T_C\simeq0.1$ \cite{Tinkham}). The resulting kinetic inductance per square of the superconducting ring is about 27.5 pH/square. We notice that the kinetic inductance is the main contribution to the total inductance of the ring.

Finally, we extracted the resistance of a single JJ with two methods. First, by analyzing the $I-V$ characteristics in Fig. \ref{fig:Fig1}-c and considering that the measured normal-state resistance is given from the parallel connection of two Dayem bridges, we get a resistance for each junction $R_{JJ}\simeq10$ $\Omega$. Second, we evaluated the normal-state resistance of each constriction from the conductance of our thin film and the geometry of the JJs. In particular we got $R_{JJ}=l/\sigma wt\simeq9.6$ $\Omega$, where $l=150$ nm is the junction length, $w=150$ nm is the constriction width and $t=30$ nm is the thickness of the thin film. It follows that the two methods provide results in good agreement.

\subsection*{Theoretical model}
The operation of a SQUID based on Dayem bridges can be determined by solving the system of resistively and capacitively shunted junction (RCSJ) equations~\cite{Gra16}:
\begin{equation}\label{RCSJa}
\frac{I}{2}+I_{circ}+I_L^{\text{th}}+I_L^{\text{gate}}=\frac{\hbar }{2e}C_L\ddot{\varphi}_L+\frac{\hbar}{2e}\frac{1}{R_L}\dot{\varphi}_L+I_L(\varphi_L)
\end{equation}
\begin{equation}\label{RCSJb}
\frac{I}{2}-I_{circ}+I_R^{\text{th}}+I_R^{\text{gate}}=\frac{\hbar }{2e}C_R\ddot{\varphi}_{R}+\frac{\hbar}{2e}\frac{1}{R_R}\dot{\varphi}_R+I_R(\varphi_R),
\end{equation}
where $I$ is the external bias current, $I_{circ}$ is the current circulating in the superconducting ring, and $e$ is the electron charge. Moreover, $C_i$, $R_i$, $I_{i}$, and $\varphi_i$ are the capacitance, the normal-state resistance, the Josephson current, and the phase difference of the $i$-th JJ (with $i=L,R$), respectively.
In Eqs.~\eqref{RCSJa}-\eqref{RCSJb}, we included the Johnson–Nyquist thermal noise terms, $I_i^{\text{th}}$, and the gate-dependent Gaussianly-distributed, delta-correlated stochastic noise fluctuations, $I_i^{\text{gate}}$, with intensity $D^{\text{gate}}_i$.

The critical temperature of our titanium thin film is $T_C\simeq420$ mK, therefore we are in the temperature regime ($T=50$ mK $< 0.4T_C$) where the superconducting gap shows its zero-temperature value $\Delta_i$ \cite{Tinkham}. In addition, in our experiment we have $l/\xi\simeq1.5$, so that, since the short junction equations are fairly valid for ($l/\xi<2.5$)~\cite{Virtanen2016}, we approximate the CPR of the JJs with the zero-temperature non-sinusoidal Kulik-Omel'Yanchuk CPR for diffusive short junctions KO-1~\cite{Kul75}:
\begin{equation}
I_i(\varphi_i)=\frac{\pi\Delta_i}{eR_i}\cos\left ( \frac{\varphi_i}{2} \right )\tanh ^{-1}\left [ \sin\left ( \frac{\varphi_i}{2} \right ) \right ].
\label{CPRDaymeBridges}
\end{equation}
Then, the critical current of each JJ is given by:
\begin{equation}
I_{C,i}={\max}_{\varphi}I_i(\varphi)\simeq\frac{2\pi\Delta_i}{3eR_i}.
\label{CPRDaymeBridges}
\end{equation}

The numerical solution of Eqs.~\eqref{RCSJa}-\eqref{RCSJb} demands the imposition of flux quantization in a superconducting ring interspersed with two weak links, that can be expressed as:
\begin{equation}
\varphi_L-\varphi_R=2\pi\frac{\Phi}{\Phi_0} -\beta \frac{I_{circ}}{\overline{I}_C}+2\pi k.
\label{Fluxquantization}
\end{equation}
Here, $\Phi$ is the external magnetic flux, $k$ is the integer number of enclosed flux quanta in the ring, $\overline{I}_C=(I_{C,L}+I_{C,R})/2$, while $\beta$ is the screening parameter defined as
\begin{equation}
\beta=\frac{2\pi \mathcal{L}}{\Phi_0}\overline{I}_C,
\label{ScreeningParameter}
\end{equation}
with $\mathcal{L}$ being the total inductance of the superconducting ring.

Further details about the assumptions made and the values imposed for the system parameters, i.e., for the critical current, the inductance of SQUID arm, and the amplitude $D^{\text{gate}}_i$ of the additional noise source, in order to well reproduce the experimental data, are given in the Supplementary materials.

\section*{Associated Content}
Supplementary Information available.

\section{Author Contributions}
F.P. fabricated the samples. F.P., F.V. and G.D.S. performed the measurements. F.P. and F.V. analysed the experimental data with input from G.D.S. and F.G.. C.G. developed the theoretical
model with inputs from F.G. and P.S.. G.D.S. performed the finite element simulations. F.G. conceived the experiment. F.P. wrote the manuscript with input from all authors. All authors discussed the results and their implications equally at all stages.

\section{Notes}
The authors declare no competing financial interest.

\section*{Acknowledgement}
We acknowledge F. S. Bergeret, A. Braggio, V. Golovach, E. Strambini, S. Kafanov, A. Romito and Y. Pashkin for fruitful discussions.
The authors acknowledge the European Research Council under the European Union's Seventh Framework Programme (COMANCHE; European Research Council Grant No. 615187) and
Horizon 2020 and innovation programme under grant agreement No. 800923-SUPERTED. The work of G.D.S. and F.P. was partially funded by the Tuscany Region under the FARFAS 2014
project SCIADRO. The work of F.P. was partially supported by the Tuscany Government (Grant No. POR FSE 2014-2020) through the INFN-RT2 172800 project.


\begin{thebibliography}{99}

\bibitem{DeSimoni2018}
De Simoni, G.; Paolucci, F.; Solinas, P.; Strambini, E.; Giazotto, F.
Metallic supercurrent field-effect transistor,
\textit{Nat. Nanotech.} {\bf{2018}}, 13, 802-805.

\bibitem{Varnava2018}
Varnava, C.
Transistors go metal,
\textit{Nat. Electron.} {\bf{2018}}, 1, 374.

\bibitem{Paolucci2018}
Paolucci, F.; De Simoni, G.; Strambini, E.; Solinas, P.; Giazotto, F.
Ultra-Efficient Superconducting Dayem Bridge Field-Effect Transistor,
\textit{Nano Lett.} {\bf{2018}}, 18, 4195-4199.

\bibitem{Paolucci2019}
Paolucci, F.; De Simoni, G.; Solinas, P.; Strambini, E.; Ligato, N.; Virtanen, P.; Braggio, A.; Giazotto, F.
Magnetotransport Experiments on Fully Metallic Superconducting Dayem-Bridge Field-Effect Transistors,
\textit{Phys. Rev. Appl.} {\bf{2019}}, 11, 024061.

\bibitem{DeSimoni2019}
De Simoni, G.; Paolucci, F.; Puglia, C.; Giazotto, F.
Mesoscopic Al/Cu/Al Josephson field-effect transistors,
\textit{ACS Nano} {\bf{2019}}, 13, 7871-7876.

\bibitem{Virtanen2019}
Virtanen, P.; Braggio, A.; Giazotto, F.
Superconducting size effect in thin films under electric field: mean-field self-consistent model,
\textit{arXiv:1903.01155} {\bf{2019}}.

\bibitem{vanDam2006}
van Dam, J. A.; Nazarov Y. V.; Bakkers, E. P. A. M.; De Franceschi, S.; Kouwenhoven, L. P.
Supercurrent reversal in quantum dots,
\textit{Nature} {\bf{2006}}, 442, 667-670.

\bibitem{Cleuziou2006}
Cleuziou, J.-P.; Wernsdorfer, W.; Bouchiat, V.; Ondar�uhu, T.; Monthioux, M.
Carbon nanotube superconducting quantum interference device,
\textit{Nat. Nanotech.} {\bf{2006}}, 1, 53-59.

\bibitem{Girit2008}
Girit, C.; Bouchiat, V.; Naaman, O.; Zhang, Y.; Crommie, M. F.; Zettl, A.; Siddiqi, I.
Tunable Graphene dc Superconducting Quantum Interference Device,
\textit{Nano Lett.} {\bf{2008}}, 9, 198-199.

\bibitem{Goswami2016}
Goswami, S.; Mulazimoglu, E.; Monteiro, A. M. R. V. L.; W\"olbing, R.; Koelle, D.; Kleiner, R.; Blanter, Ya. M.; Vandersypen, Ya. M.; Caviglia, A. D.
Quantum interference in an interfacial superconductor,
\textit{Nat. Nanotech.} {\bf{2016}}, 11, 861-865.

\bibitem{Clarke2004}
Clarke, J.; Braginski, J. \textit{The SQUID Handbook} VCH: New York, 2004.

\bibitem{Clark1980}
Clark, T. D.; Prance, R. J.; Grassie, A. D. C.
Feasibility of hybrid Josephson field effect transistors,
\textit{J. Appl. Phys.} {\bf{1980}}, 51, 2736-2743.

\bibitem{Terzioglu1998}
Terzioglu, E.; Beasley, M.~R.
Complementary Josephson junction devices and circuits: a possible new approach to superconducting electronics,
\textit{IEEE Trans. Appl. Supercond.} {\bf{1998}}, 8, 48-53.

\bibitem{Devoret2013}
Devoret, M.~H.; Schoelkopf, R.~J.
Superconducting Circuits for Quantum Information: An Outlook,
\textit{Science} {\bf{2013}}, 339, 1169-1174.

\bibitem{Majer2002}
Majer, J.~B.; Butcher, J.~R.; Mooij, J.~E.
Simple phase bias for superconducting circuits,
\textit{Appl. Phys. Lett.} {\bf{2002}}, 80, 3638-3640.

\bibitem{Ioffe1999}
Ioffe, L.~B.; Geshkenbein, V.~B.; Fiegel'man, M.~V.; Fauchere, A.~L.; Blatter, G.
Environmentally decoupled sds -wave Josephson junctions for quantum computing,
\textit{Nature} {\bf{1999}}, 398, 679-681.

\bibitem{Mooij1999}
Mooij, J.~E.; Orlando, T.~P.;  Levitov, L.;  Tian, L.; van der Wal, C.~H.; Lloyd, S.
Josephson Persistent-Current Qubit,
\textit{Science} {\bf{1999}}, 285, 1036-1039.

\bibitem{Anahory2014}
Anahory, Y.; Reiner, J.; Embon, L.; Halbertal, D.; Yakovenko, A.; Myasoedov, Y.; Rappaport, M.~L.; Huber, M.~E.; Zeldov, E.
Three-Junction SQUID-on-Tip with Tunable In-Plane and Out-of-Plane Magnetic Field Sensitivity,
\textit{Nano Lett.} {\bf{2014}}, 14, 6481-6487.

\bibitem{Uri2016}
Uri, A.; Meltzer, A.~Y.; Anahory, Y.; Embon, L.; Lachman, E.~O.; Halbertal, D.; Naren, HR.; Myasoedov, Y.; Huber, M.~E.; Young, A.~F.; Zeldov, E.
Electrically Tunable Multiterminal SQUID-on-Tip,
\textit{Nano Lett.} {\bf{2016}}, 16, 6910-6915.

\bibitem{Fulton1972}
Fulton, T.~A.; Dunkleberger, L.~N.;; Dynes, R.~C.
Quantum Interference Properties of Double Josephson Junctions,
\textit{Phys. Rev. B.} {\bf{1972}}, 6, 855-875.

\bibitem{Tsang1975}
Tsang, W.-T.; Van Duzer, T
dc analysis of parallel arrays of two and three Josephson junctions,
\textit{J. Appl. Phys.} {\bf{1975}}, 46, 4573-4580.

\bibitem{Russo2014}
Russo, R.; Granata, C.; Vettoliere, A.; Esposito, E.; Fretto, M.; De Leo, N.; Enrico, E.; Lacquaniti, V.
Performances of niobium planar nanointerferometers as a function of the temperature: a comparative study,
\textit{Supercond. Sci. Technol.} {\bf{2014}}, 27, 044028.

\bibitem{Bar82}
Barone, A.; Patern\`{o}, G.
\textit{Physics and Applications of the Josephson Effect} (Wiley: New York, 1982).

\bibitem{Morpurgo1998}
Morpurgo, A.; Klapwijk, T.~M.; van Wees, B.~J.
Hot electron tunable supercurrent. \textit{Appl. Phys. Lett.} {\bf{1998}}, 72, 966-968.

\bibitem{Tinkham}
Tinkham, M. \emph{Introduction to Superconductivity} (McGraw-Hill: New York, 1996).

\bibitem{Gra16}
Granata, C.; Vettoliere, A.
Nano Superconducting Quantum Interference device: A powerful tool for nanoscale investigations,
\textit{Phys. Rep.} {\bf{2016}}, 614, 1-69.

\bibitem{Kul75}
Kulik, I.; Omel'Yanchuk, A.
Contribution to the microscopic theory of the Josephson effect in superconducting bridges,
\textit{Sov. J. Exp. Theor. Phys. Lett.} {\bf{1975}}, 21, 96-97.

\bibitem{Sol15}
Solinas, P.; Gasparinetti, S.; Golubev, D.; Giazotto, F.
A Josephson radiation comb generator,
\textit{Sci. Rep.} {\bf{2015}}, 5, 12260.

\bibitem{GuaSol18}
Guarcello, C.; Solinas, P.; Braggio, A.; Di Ventra, M.; Giazotto, F.
Josephson Thermal Memory,
\textit{Phys. Rev. Applied} {\bf{2018}}, 9, 014021.

\bibitem{Sleator1995}
Sleator, T.; Weinfurter, H.
Realizable Universal Quantum Logic Gates,
\textit{Phys. Rev. Lett.} {\bf{1995}}, 74, 4087-4090.

\bibitem{Lloyd1995}
Lloyd, S.
Almost Any Quantum Logic Gate is Universal,
\textit{Phys. Rev. Lett.} {\bf{1995}}, 75, 346-349.

\bibitem{Yan2016}
Yan, F.; Gustavsson, S.; Kamal, A.; Birenbaum, J.; Sears, A. P.; Hover, D.; Gudmundsen, T. J.; Rosenberg, D.; Samach, G.; Weber, S.; Yoder, J. L.; Orlando, T. P.; Clarke, J.; Kerman, A. J.; Oliver, W. D.
The flux qubit revisited to enhance coherence and reproducibility,
\textit{Nat. Commun.} {\bf{2016}}, 7, 12964.

\bibitem{Martinis2002}
Martinis, J.~M.; Nam, S.; Aumentado, J.; Urbina, C.
Rabi Oscillations in a Large Josephson-Junction Qubit,
\textit{Phys. Rev. Lett.} {\bf{2002}}, 89, 117901.

\bibitem{Koch2007}
Koch, J.; Yu, T. M.; Gambetta, J.; Houck, A. A.; Schuster, D. I.; Majer, J.; Blais, A.; Devoret, M. H.; Girvin. S. M.; Schoelkopf R. J.
Charge-insensitive qubit design derived from the Cooper pair box,
\textit{Phys. Rev. A} {\bf{2007}}, 76, 042319.

\bibitem{Schreier2008}
Schreier, J. A.; Houck, A. A.; Koch, J.; Schuster, D. I.; Johnson, B. R.; Chow, J. M.; Gambetta, J. M.; Majer, J.; Frunzio, L.; Devoret, M. H.; Girvin. S. M.; Schoelkopf R. J.
Suppressing charge noise decoherence in superconducting charge qubits,
\textit{Phys. Rev. B} {\bf{2008}}, 77, 180502(R).

\bibitem{Barends2013}
Barends, R.; Kelly, J.; Megrant, A.; Sank, D.; Jeffrey, E.; Chen, Y.; Yin, Y.; Chiaro, B.; Mutus, J.; Neill, C.; O�Malley,P.; Roushan, P.; Wenner, J.; White, T. C.; Cleland, A. N.; Martinis J. M.
Coherent Josephson Qubit Suitable for Scalable Quantum Integrated Circuits,
\textit{Phys. Rev. Lett.} {\bf{2013}}, 111, 080502.

\bibitem{Likharev1991}
Likharev, K.~K.; Semenov, V.~K.
RSFQ logic/memory family: a new Josephson-junction technology for sub-terahertz-clock-frequency digital systems,
\textit{IEEE Trans. Appl. Supercond.} {\bf{1991}}, 1, 3-28.

\bibitem{Worsham1995}
Worsham, A. H.; Przybysz, J. X.; Kang, J.; Miller, D. L.
A Single Flux Quantum cross-bar switch and demultiplexer,
\textit{IEEE Trans. Appl. Supercond.} {\bf{1995}}, 5, 2996-2999.

\bibitem{Fornieri2017}
Fornieri, A.; Giazotto, F.
Towards phase-coherent caloritronics in superconducting circuits,
\textit{Nat. Nanotech.} {\bf{2017}}, 12, 944-952.

\bibitem{Giazotto2014}
Giazotto, F.; Robinson, J. W. A.; Moodera, J. S.; Bergeret, F. S.
Proposal for a phase-coherent thermoelectric transistor,
\textit{Appl. Phys. Lett} \textbf{2014}, 105, 062602.

\bibitem{Virtanen2018}
Virtanen, P.; Ronzani, A.; Giazotto, F.
Josephson Photodetectors via Temperature-to-Phase Conversion,
\textit{Phys. Rev. Applied} {\bf{2018}}, 9, 054027.

\bibitem{Hurley1995}
Hurley, W. G:; Duffy, M. C.
Calculation of self and mutual impedances in planar magnetic structures,
\textit{IEEE Trans. Magn.} {\bf{1995}}, 31, 2416-2422.

\bibitem{Mohan1999}
Mohan, S. S.; del Mar Hershenson, M.; Boyd, S. P.; Lee, T. H.
Simple accurate expressions for planar spiral inductances,
\textit{IEEE J. Solid-State Circuits} {\bf{1999}}, 34, 1419-1424.

\bibitem{Annunziata2010}
Annunziata, A. J.; Santavicca, D. F.; Frunzio, L.; Catelani, G.; Rooks, M. J.
Tunable superconducting nanoinductors,
\textit{Nanotechnology} {\bf{2010}}, 21, 445202.

\bibitem{Virtanen2016}
Virtanen, P.; Ronzani, A.; Giazotto, F.
Spectral Characteristics of a Fully Superconducting SQUIPT,
\textit{Phys. Rev. Applied} {\bf{2016}}, 6, 054002.


\end{thebibliography}
\end{document}